\documentstyle[12pt]{article} 
\input psfig.sty

\def\Title#1{\begin{center} {\Large #1 } \end{center}}
\def\Author#1{\begin{center}{ \sc #1} \end{center}}
\def\Address#1{\begin{center}{ \it #1} \end{center}}

\def\doeack{\footnote{Work supported by the Department of Energy,
                     contract DE--AC03--76SF00515.}}
\def\SLAC{Stanford Linear Accelerator Center\\
    Stanford University, Stanford, California 94309 USA}

\newenvironment{Abstract}{\begin{quotation} \begin{center}
                       ABSTRACT
     \end{center}\bigskip  }{\end{quotation}}
\def\beq{\begin{equation}}
\def\eeq#1{\label{#1}\end{equation}}
\def\eeqn{\end{equation}}
\def\beqa{\begin{eqnarray}}
\def\eeqa#1{\label{#1}\end{eqnarray}}
\def\eeqan{\end{eqnarray}}

\def\Acknowledgements{\bigskip  \bigskip \begin{center} \begin{large}
             \bf ACKNOWLEDGEMENTS \end{large}\end{center}}

\def\Re{{\cal R \mskip-4mu \lower.1ex \hbox{\it e}\,}}
\def\Im{{\cal I \mskip-5mu \lower.1ex \hbox{\it m}\,}}
\def\nn{\noindent}
\def\ie{{\it i.e.}}
\def\eg{{\it e.g.}}

\def\etal{{\it et al.}}
\def\ibid{{\it ibid}.}
\def\sub#1{_{\lower.25ex\hbox{$\scriptstyle#1$}}}
\def\sul#1{_{\kern-.1em#1}}
\def\sll#1{_{\kern-.2em#1}}  
\def\sbl#1{_{\kern-.1em\lower.25ex\hbox{$\scriptstyle#1$}}}
\def\ssb#1{_{\lower.25ex\hbox{$\scriptscriptstyle#1$}}}
\def\sbb#1{_{\lower.4ex\hbox{$\scriptstyle#1$}}}

\def\to{\rightarrow}
\def\dk{\ifmmode \Delta\kappa\else $\Delta\kappa$\fi}
\def\sigt{\ifmmode \tilde\sigma\else $\tilde\sigma$\fi}
\def\mh{\ifmmode m\sbl H \else $m\sbl H$\fi}
\def\mch{\ifmmode m_{H^\pm} \else $m_{H^\pm}$\fi}
\def\mt{\ifmmode m_t\else $m_t$\fi}
\def\mc{\ifmmode m_c\else $m_c$\fi}
\def\mz{\ifmmode M_Z\else $M_Z$\fi}
\def\mw{\ifmmode M_W\else $M_W$\fi}
\def\mws{\ifmmode M_W^2 \else $M_W^2$\fi}
\def\mhs{\ifmmode m_H^2 \else $m_H^2$\fi}   
\def\mzs{\ifmmode M_Z^2 \else $M_Z^2$\fi}
\def\mts{\ifmmode m_t^2 \else $m_t^2$\fi}
\def\mcs{\ifmmode m_c^2 \else $m_c^2$\fi}
\def\mchs{\ifmmode m_{H^\pm}^2 \else $m_{H^\pm}^2$\fi}
\def\ztwo{\ifmmode Z_2\else $Z_2$\fi}
\def\zone{\ifmmode Z_1\else $Z_1$\fi}
\def\mtwo{\ifmmode M_2\else $M_2$\fi}
\def\mone{\ifmmode M_1\else $M_1$\fi}
\def\tb{\ifmmode \tan\beta \else $\tan\beta$\fi}
\def\xw{\ifmmode x\sub w\else $x\sub w$\fi}
\def\ch{\ifmmode H^\pm \else $H^\pm$\fi}
\def\lum{\ifmmode {\cal L}\else ${\cal L}$\fi}
\def\inpb{\ifmmode {\rm pb}^{-1}\else ${\rm pb}^{-1}$\fi}
\def\infb{\ifmmode {\rm fb}^{-1}\else ${\rm fb}^{-1}$\fi}
\def\epem{\ifmmode e^+e^-\else $e^+e^-$\fi}
\def\ppb{\ifmmode \bar pp\else $\bar pp$\fi}

\def\bsg{\ifmmode b\rightarrow s\gamma \else $b\rightarrow s\gamma$\fi}
 
\newskip\zatskip \zatskip=0pt plus0pt minus0pt
\def\matth{\mathsurround=0pt}

\def\atversim#1#2{\lower0.7ex\vbox{\baselineskip\zatskip\lineskip\zatskip
  \lineskiplimit 0pt\ialign{$\matth#1\hfil##\hfil$\crcr#2\crcr\sim\crcr}}}

\begin{document}
\rightline{\vbox{\halign{&#\hfil\cr
&SLAC-PUB-7317\cr
&October 1996\cr}}}
\vspace{0.8in} 
\Title{Using Final State Gluons as Probes of Anomalous Top Quark Couplings 
at the NLC
}
\bigskip
\Author{Thomas G. Rizzo\doeack}
\Address{\SLAC}
\bigskip
\begin{Abstract}
 
The rate and corresponding gluon jet energy distribution for the process 
$e^+e^- \to t\bar tg$ are sensitive to the presence of anomalous dipole-like 
couplings of the top to the photon and $Z$ at the production vertex as well 
as to the gluon itself. For sizeable anomalous couplings substantial 
deviations in the shape and magnitude of the gluon spectrum from the 
expectations of the Standard Model are anticipated. We explore the capability 
of the Next Linear Collider to either discover or place bounds on these types 
of top quark couplings through measurements of the gluon 
energy distribution. The resulting constraints are found to be quite 
complementary to those obtained using other techniques. 

\end{Abstract}
\bigskip
\vskip1.0in
\begin{center}
To appear in the {\it Proceedings of the 1996 DPF/DPB Summer Study on New
 Directions for High Energy Physics-Snowmass96}, Snowmass, CO, 
25 June-12 July, 1996. 
\end{center}
%
\bigskip
\def\thefootnote{\fnsymbol{footnote}}
\setcounter{footnote}{0}
\newpage
\section{Introduction}

The Standard Model(SM) has provided a remarkably successful description of 
almost all available data involving the strong and electroweak interactions. 
In particular, the discovery of the top quark at the Tevatron with a 
mass{\cite {tev}}, $m_t=175\pm 6$ GeV, close to that anticipated by fits to 
precision electroweak data, is indeed a great triumph. However, 
the fact that both $R_b$ and $A_b$ remain{\cite {moriond}} approximately 
$2\sigma$ away  
from SM expectations may be providing us with the first indirect window 
into new physics. In fact, this apparent deviation in $b$-quark couplings 
from the SM expectations allows one to speculate that perhaps all of the 
members of the third generation might couple to some kind of new physics. 
Independent of these potential discrepancies with the SM, 
since the top is the most massive fermion, it is believed by many 
that the detailed physics of the top quark may be significantly different 
than what is predicted by the SM. This suggestion makes precision measurements 
of all of the top quark's properties mandatory. 

Perhaps the most obvious and easily imagined scenario is one in which 
the top's 
couplings to the SM gauge bosons, \ie, the $W$, $Z$, $\gamma$, and $g$, 
are altered. This possibility, extended to all of the fermions of the third 
generation, has attracted a lot of attention over the last few 
years{\cite {big}}. In the case of the electroweak interactions involved in 
top pair production in $e^+e^-$ collisions, the lowest dimensional 
gauge-invariant, non-renormalizable operators representing new physics that 
we can introduce take 
the form of dipole moment-type couplings to the $\gamma$ and $Z$. The 
anomalous magnetic moment operators, 
which we can parameterize by a pair of dimensionless quantities, 
$\kappa_{\gamma,Z}$, are $CP$-conserving. The corresponding electric dipole 
moment terms, parameterized as $\tilde \kappa_{\gamma,Z}$, are $CP$-violating. 
Clearly, analogous interactions may also be anticipated for the top couplings 
with gluons which is directly involved in top production at hadron colliders. 
The shift in the three-point $t\bar t\gamma,Z,g$ interactions due 
to the existence of these anomalous couplings can be written as 
\begin{equation}
{\delta \cal L}={i\over {2m_t}}\bar t
\sigma_{\mu\nu}q^{\nu}\left[g_sT_a(\kappa_g^t-i\tilde\kappa_g
^t\gamma_5)G^{a\mu}+e(\kappa_\gamma^t-i\tilde\kappa_
\gamma^t\gamma_5)A^{\mu}+{g\over {2c_w}}(\kappa_Z^t-i\tilde\kappa_Z^t
\gamma_5)Z^{\mu}\right]t \,,
\end{equation}
where $e$ is the proton charge, $g(g_s)$ is the standard weak(strong) 
coupling constant, $T_a$ are the color generators,
$c_w=\cos \theta_W$, and $q$ is the $\gamma,g$ or $Z$'s four-momentum. Gauge 
invariance will also lead to new four-point interactions involving two  
gauge bosons and the top, \eg, $t\bar t gg$ and $t\bar t W^+W^-$,  
but they will not concern us here as we will only work to leading order in the 
strong and electroweak interactions. In most cases gauge invariance will 
relate any of the trilinear $t\bar tZ,\gamma$ anomalous couplings to others 
involving the 
$tbW$ vertex. Escribano and Masso{\cite {big}} have shown that 
{\it in general} all 
of the anomalous three-point couplings involving the neutral gauge bosons can  
be unrelated even when the underlying operators are SM gauge invariant. Thus in 
our analysis we will treat all $\kappa$'s and $\tilde \kappa$'s 
as independent free parameters. Of course, within any  
particular new physics scenario the anomalous couplings will no longer be 
completely independent. For example, if the new physics generated an effective 
dimension-6 operator which coupled to top via its hypercharge quantum number, 
we would find the relation
\begin{equation}
\kappa^t_\gamma={-\kappa^t_Z \over {2sin^2 \theta_w}} \,,
\end{equation}
between the photon and $Z$ anomalous magnetic dipole couplings with an 
identical expression holding for the electric dipole couplings. (A similar 
but somewhat different result occurs if the new operator coupled instead to 
the top quark's weak isospin in 
which case $\sin^2 \theta_w \to -\cos^2 \theta_w$ in the expression above.)

As has been discussed in the literature{\cite {big}}, if any of 
the anomalous couplings are sufficiently large their effects can be directly 
probed by top pair production at either $e^+e^-$ or hadron colliders. The 
purpose of the present work is to consider 
the sensitivity of the process $e^+e^- \to t\bar tg$ to non-zero values of any 
of these anomalous couplings.

\section{Analysis I: Chromoelectric and Chromomagnetic Moments}

The basic cross section formulae and analysis procedure for either anomalous 
strong or electroweak dipole couplings can be found 
in Ref.{\cite {old}}. The essential observation of our analysis is that the 
presence of anomalous top couplings {\it at either vertex} modifies the 
shape of the gluon energy 
spectrum in the $e^+e^- \to t\bar tg$ process. An example of this is shown in 
Fig.~1 in the case of anomalous $t\bar tg$ couplings for $\sqrt s=500$ and 1000 
GeV. (In this figure, $z=2E_g/\sqrt {s}$, where $E_g$ is the gluon jet 
energy.) Here, we will go beyond the 
initial studies that exist in the literature in several 
ways; in the case of anomalous chromomagnetic and chromoelectric we have done 
the following: 
($i$) we generalize the form of the $t\bar tg$ coupling to allow for 
the possibility of a 
sizeable chromoelectric moment, $\tilde \kappa$. The incorporation of 
$\tilde \kappa \neq 0$ into the expressions for the differential cross section 
in Ref.{\cite {old}} is rather 
straightforward and can be accomplished by the simple substitution 
$\kappa^2 \to \kappa^2+\tilde \kappa^2$ made universally. 
Note that since a non-zero value of $\tilde \kappa$ 
produces a $CP$-violating interaction it appears only quadratically in the 
expression for the gluon energy distribution since this is a $CP$-conserving 
observable. Thus, in comparison 
to $\kappa$, we anticipate a greatly reduced sensitivity to the value of 
$\tilde \kappa$. ($ii$) We use updated 
expectations for the available integrated luminosities of the NLC at various 
center of mass energies as well as an updated efficiency($\simeq 100\%$) for 
identifying top-quark pair production events. Both of these changes obviously 
leads to a direct increase in statistical power compared to 
Ref.{\cite {old}} ($iii$) Perhaps, even more importantly, we soften the 
cut placed 
on the minimum gluon jet energy, $E_g^{min}$, in performing the energy spectrum 
fits.

\vspace*{-0.5cm}
\nn
\begin{figure}[htbp]
\centerline{
\psfig{figure=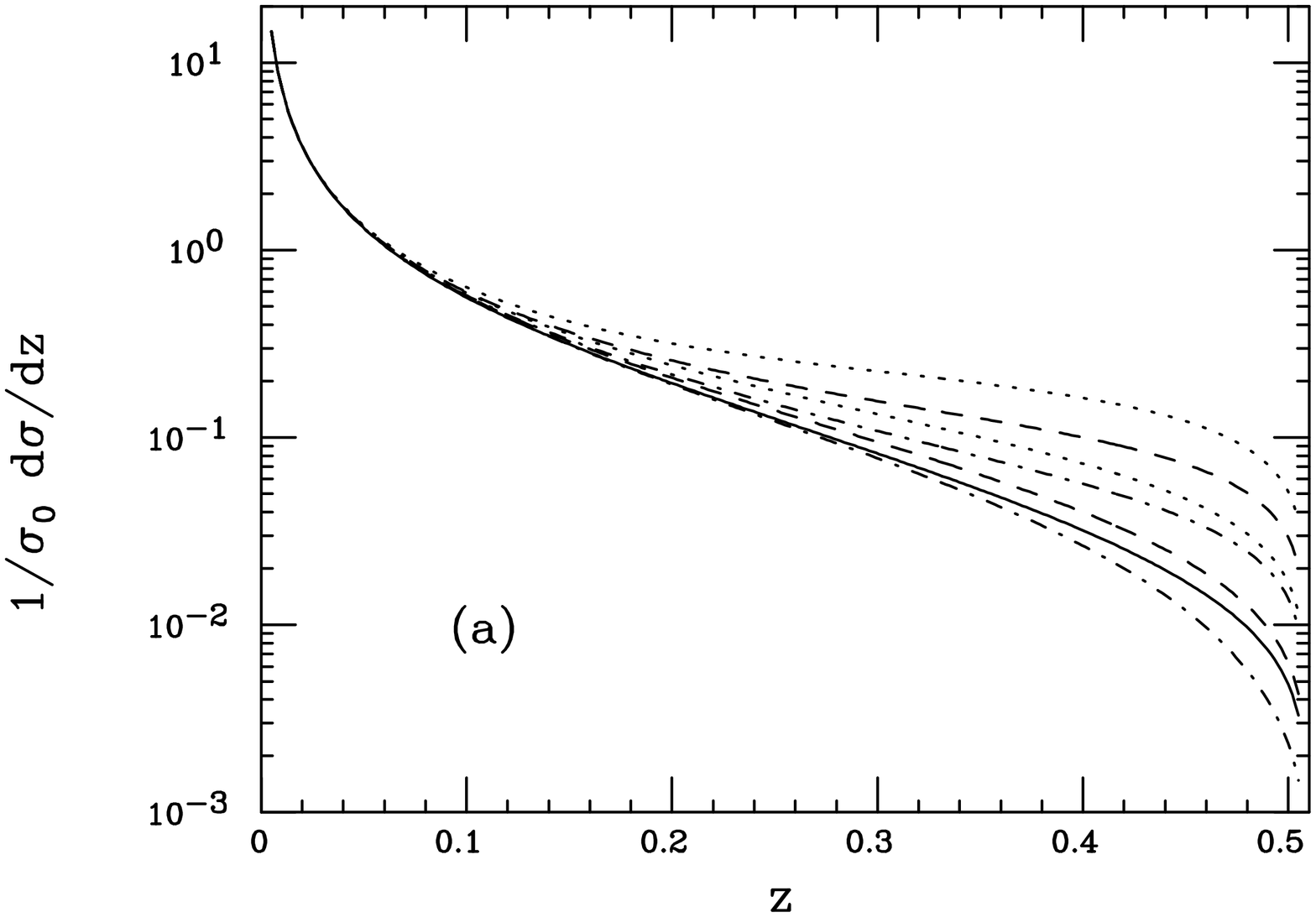,height=9.1cm,width=9.1cm,angle=-90}
\hspace*{-5mm}
\psfig{figure=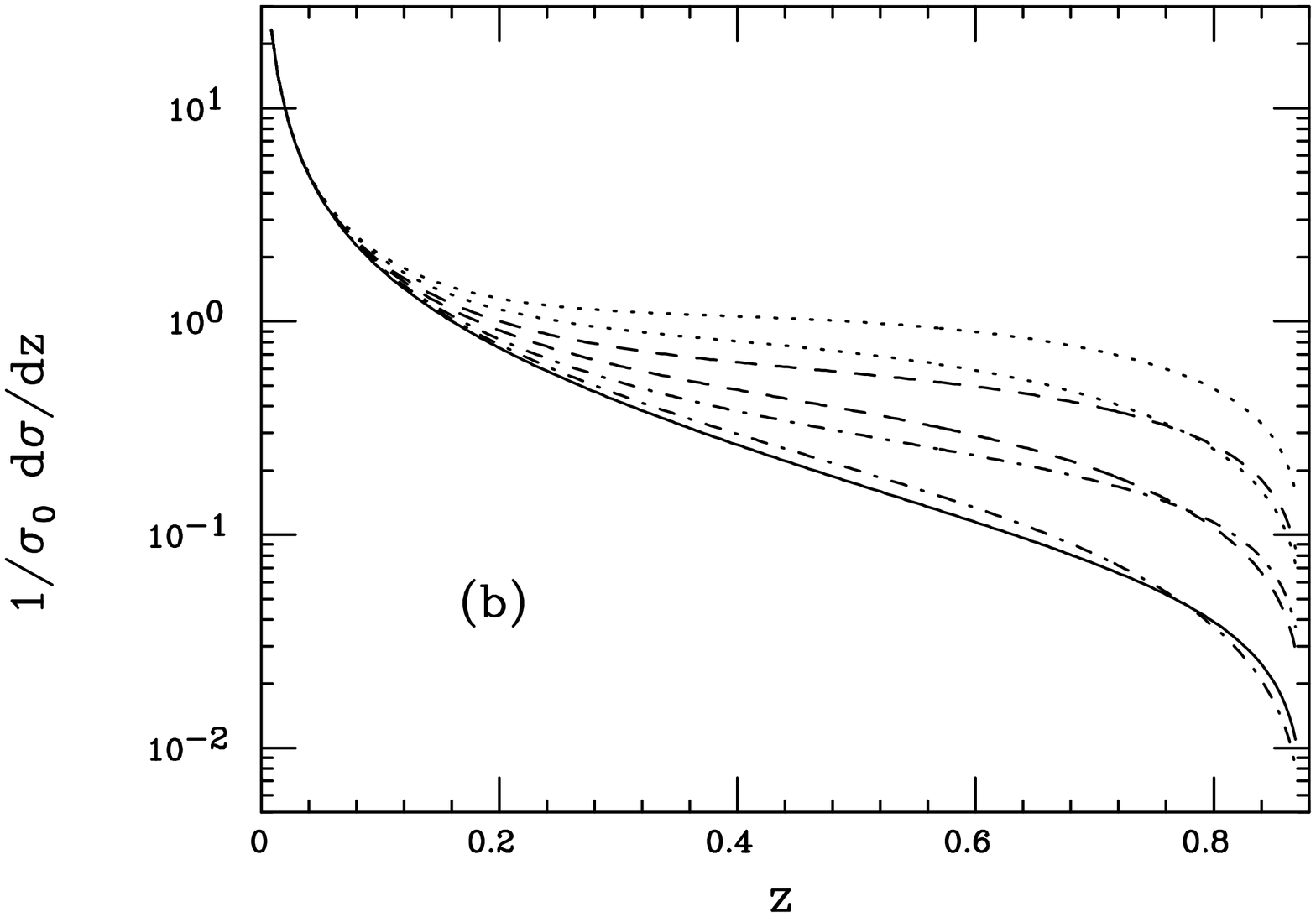,height=9.1cm,width=9.1cm,angle=-90}}
\vspace*{-0.6cm}
\caption{Gluon jet energy spectrum assuming $\alpha_s=0.10$ for $m_t=175$ 
GeV at a center of mass energy of (a) 500 GeV or (b) 1 TeV NLC. The 
upper(lower) dotted, 
dashed, and dot-dashed curves correspond to $\kappa$ values of 3(-3), 2(-2), 
and 1(-1) respectively while the solid curve is conventional QCD with 
$\kappa=0$.}
\end{figure}
\vspace*{0.1mm}

The reasons for having such a cut are two-fold. First, a minimum gluon 
energy is required to identify the event as $t\bar tg$. The cross section 
itself is infra-red singular 
though free of co-linear singularities due to the finite top quark mass. 
Second, since the top decays rather quickly, 
$\Gamma_t\simeq 1.45$ GeV, we need to worry about `contamination' from 
the additional gluon radiation 
off of the $b$-quarks in the final state. Such events can be effectively 
removed from our sample if we require that $E_g^{min}/\Gamma_t>>1$. In our 
past analysis we were overly conservative in our choices for $E_g^{min}$ in 
order to make 
this ratio as large as possible, \ie, we assumed $E_g^{min}=50(200)$ GeV for 
an NLC with a center of mass energy of 500(1000) GeV. It is now believed that 
we can with reasonable justification soften these cuts to at least as low a 
value as 
25(50) GeV for the same center of mass energies{\cite {orr}}, with a potential  
further softening of the cut at the higher energy machine being possible. 
Due to the dramatic 
infra-red behaviour of the cross section, this change in the cuts leads not 
only to an increased statistical power but also to a longer lever arm to 
probe events with very large gluon jet energies which have the most 
sensitivity to the presence of anomalous couplings. Combining all these 
modifications, as one might expect, 
we find constraints which are substantially stronger than what was obtained in 
our previous analysis{\cite {old}}.

\vspace*{-0.5cm}
\nn
\begin{figure}[htbp]
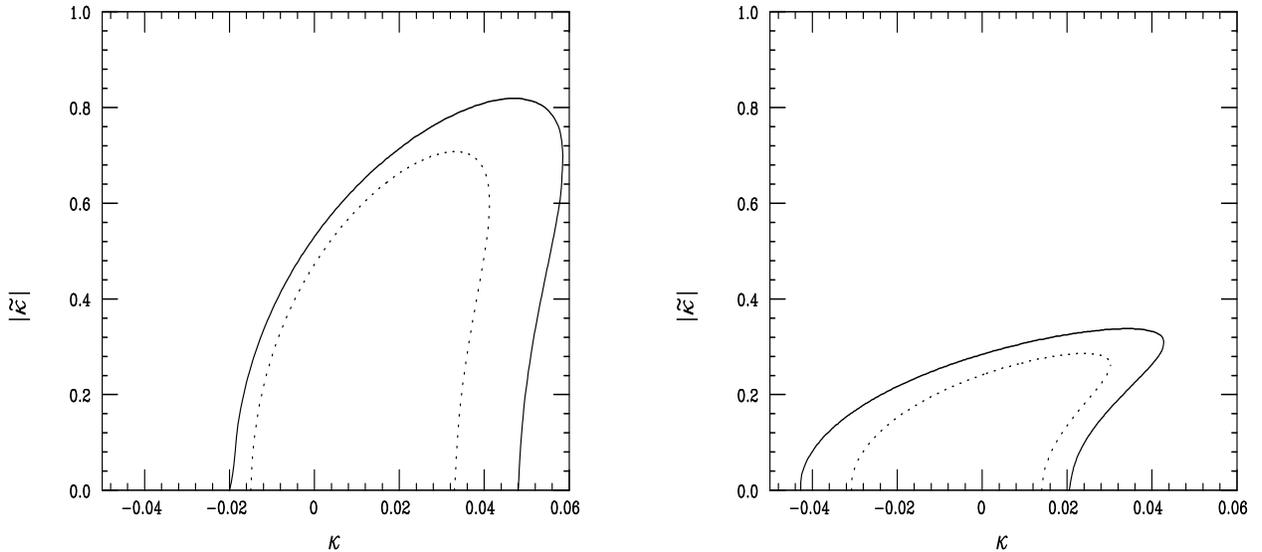

\centerline{
\psfig{figure=ttnewg.res1ps,height=9.1cm,width=9.1cm,angle=-90}
\hspace*{-5mm}
\psfig{figure=ttnewg.res2ps,height=9.1cm,width=9.1cm,angle=-90}}
\vspace*{-1cm}
\caption{\small $95\%$ CL allowed region in the $\kappa-\tilde \kappa$ plane 
obtained from fitting the gluon spectrum: on the left above $E_g^{min}$=25 
GeV at a 
500 GeV NLC assuming an integrated luminosity of 50(solid) or 100(dotted) 
$fb^{-1}$; on the right for a 1 TeV collider with $E_g^{min}$=50 
GeV and luminosities of 100(solid) and 200(dotted) $fb^{-1}$. Note that the 
allowed region has been significantly compressed downward in comparison to 
lower energy machine.}
\end{figure}
\vspace*{0.4mm}

As in Ref.{\cite {old}}, our analysis follows a Monte Carlo approach employing 
statistical errors only. For a given $e^+e^-$ center of mass energy, a binned 
gluon jet spectrum is generated for energies above $E_g^{min}$ assuming that 
the SM is correct. The bin widths are fixed to be $\Delta z=0.05$ where 
$z=2E_g/\sqrt s$ for all values of $\sqrt s$, with the number of bins thus 
determined by the values of the top mass($m_t$=175 GeV), $\sqrt s$ and 
$E_g^{min}$. As an example, at a 500 GeV 
NLC with $E_g^{min}$=25 GeV, there are 8 energy bins for the gluon energy 
spectrum beginning at $z=0.10$; 
the last bin covers the range above $z=0.45$. After the Monte Carlo data 
samples are generated, we 
perform a fit to the general expressions for the $\kappa-\tilde \kappa$ 
dependent spectrum and obtain the 
95$\%$ CL allowed region in the $\kappa-|\tilde \kappa|$ plane. Note that 
only the absolute value of $\tilde \kappa$ occurs due to reasons given above.

Fig. 2 shows the results of this procedure for a 500 GeV NLC with a cut of 
$E_g^{min}$=25 GeV for two different integrated luminosities. As expected, 
excellent constraints on $\kappa$ are now obtained but those on 
$\tilde \kappa$ are more than an order of magnitude weaker. A doubling of the 
integrated luminosity from 50 to 100 $fb^{-1}$ decreases the size of the 
allowed region by about 40$\%$. We note that in the published study 
only extremely 
poor constraints on $\kappa$ were obtained at a 500 GeV $e^+e^-$ collider, 
$-1.98\leq \kappa \leq 0.44$, due to the presence of a degenerate minima in 
the $\chi^2$ distribution. Now, with the both the increased luminosity and 
top-tagging efficiencies, as well as the longer lever arm in energy, these 
previous difficulties are circumvented.

Going to a higher energy leads to several simultaneous effects. First, since 
the cross section 
approximately scales like $\sim 1/s$ apart from phase space factors, 
a simple doubling of the 
collider energy induces a reduction in statistics unless higher integrated 
luminosities are available to compensate. Second, the sensitivity 
to the presence of 
non-zero anomalous couplings is enhanced at higher energies, roughly scaling 
like $\sim \sqrt s$ for $\kappa$ and, correspondingly, like $\sim s$ for 
$\tilde \kappa$ {\it assuming} the same available statistics at all energies. 
In Fig. 2 also we show the results of our analysis at a 1 TeV NLC 
for $E_g^{min}$=50 GeV. (Note that in our 
previous analysis, we obtained the $95\%$ CL bound $-0.12\leq \kappa \leq 0.21$ 
for this center of mass energy and an integrated luminosity of 200 
$fb^{-1}$.) For $E_g^{min}$=50 
GeV, the energy range was divided into 15 $\Delta z=0.2$ bins beginning at 
$z=0.10(0.05)$ with the last bin covering the range $z\geq 0.80$. We see from 
these figures that by going to higher energy we drastically compress the 
allowed range of $\tilde \kappa$ while the improvement for $\kappa$ is not as 
great. Lowering the energy cut is seen to lead to a far greater reduction in 
the size of the 95$\%$ CL allowed region than is a simple doubling 
of the integrated luminosity.

\section{Analysis II: Anomalous Electroweak Dipole Moments}

In this analysis we consider our observable to be the full, `normalized' 
gluon energy distribution,   
\begin{equation}
{dR\over {dz}}= {1\over {\sigma(e^+e^-\to t\bar t)}}
{d\sigma(e^+e^-\to t\bar tg)\over {dz}} \,,
\end{equation}
We note that now, unlike the previous case,  
the electroweak anomalous couplings will contribute differently 
to {\it both} the numerator and 
denominator of the expression of $dR/dz$. This implies that the sensitivity 
of $R$ to very large values(with magnitudes $\geq$ 1) of the anomalous 
couplings is quite small since their contributions effectively cancel in the 
ratio. However, for the range of anomalous couplings of interest to us 
here significant sensitivity is achieved. We follow the procedure 
as given in Ref.{\cite {old}} which also supplies the complete formulae 
for evaluating this gluon energy distribution when the $t\bar t \gamma ,Z$ 
vertex is modified by the existence of anomalous couplings. 

In comparison to our previous work, the present analysis has been extended 
in two ways. ($i$) We 
allow for the possibility that two of the four anomalous couplings may be 
simultaneously non-zero. ($ii$) As in the previous section, we lower the cut 
placed on the minimum gluon jet energy, $E_g^{min}$, in performing energy 
spectrum fits. In the published 
analysis we were again very conservative in our choices for $E_g^{min}$ in 
order to make 
this ratio as large as possible, \ie, we assumed $E_g^{min}=37.5(200)$ GeV for 
a 500(1000) GeV NLC, for the reasons described above. As in the case of the 
the anomalous strong couplings, we generate `data' following the Monte Carlo 
approach as described in the last section.

\vspace*{-0.5cm}
\nn
\begin{figure}[htbp]
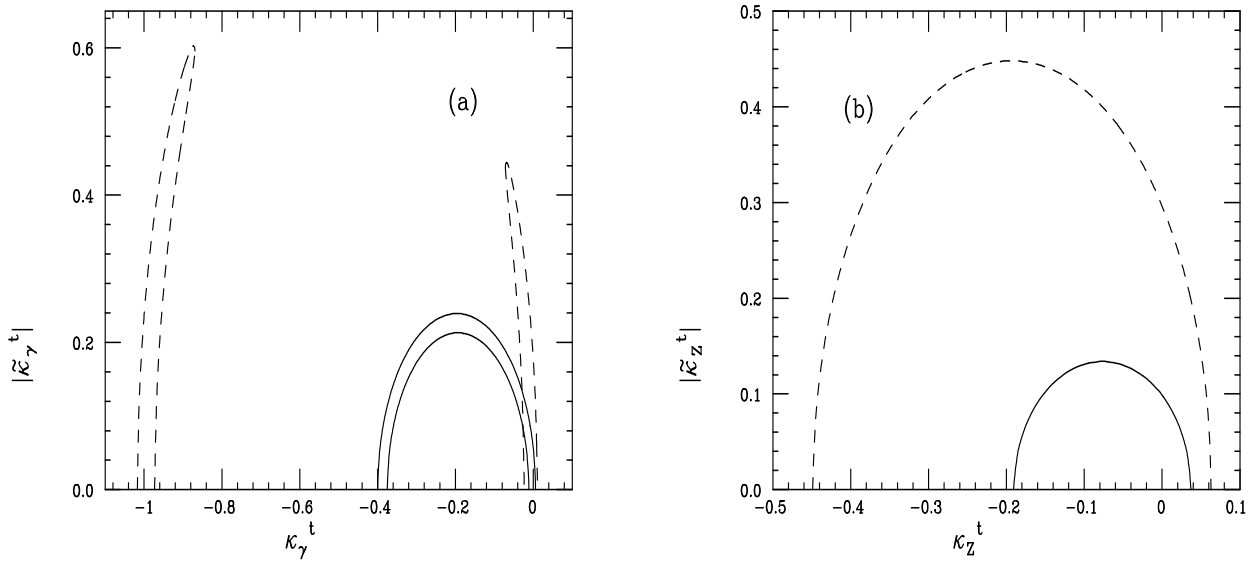

\centerline{
\psfig{figure=ttnewf.res1ps,height=9.1cm,width=9.1cm,angle=-90}
\hspace*{-5mm}
\psfig{figure=ttnewf.res2ps,height=9.1cm,width=9.1cm,angle=-90}}
\vspace*{-1cm}
\caption{\small $95\%$ CL allowed regions obtained for the anomalous couplings 
at a 500(1000) GeV NLC assuming a luminosity of 50(100) $fb^{-1}$ 
lie within the dashed(solid) curves. The gluon energy range $z\geq 0.1$ was 
used in the fit. Only two anomalous couplings are allowed 
to be non-zero at a time.}
\end{figure}
\vspace*{0.4mm}
\vspace*{-0.5cm}
\nn
\begin{figure}[htbp]
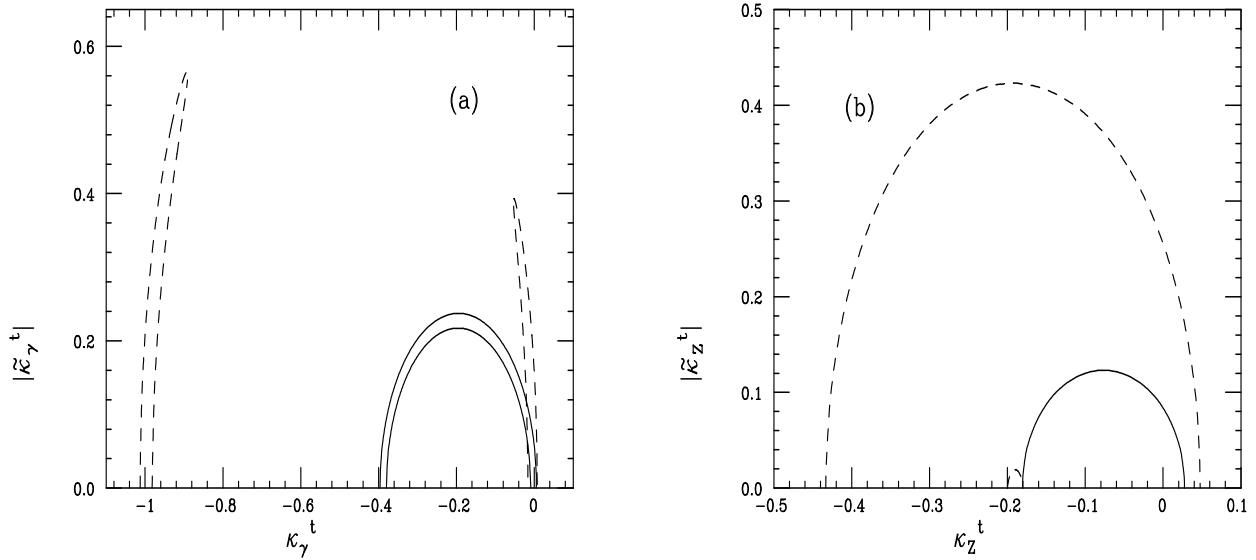

\centerline{
\psfig{figure=ttnewf.res5ps,height=9.1cm,width=9.1cm,angle=-90}
\hspace*{-5mm}
\psfig{figure=ttnewf.res8ps,height=9.1cm,width=9.1cm,angle=-90}}
\vspace*{-1cm}
\caption{\small Same as Fig. 3 but now doubling the integrated luminosity to 
100(200) $fb^{-1}$ for the 500(1000) GeV NLC.}
\end{figure}
\vspace*{0.4mm}

A fit is then performed with these data samples allowing the values of the 
various anomalous couplings to float. In general, one can perform a four 
parameter fit allowing all of the parameters $\kappa_t^{\gamma,Z}$ and 
$\tilde \kappa_t^{\gamma,Z}$ to 
be simultaneously non-zero. Here, for simplicity we allow only two of these 
couplings to be simultaneously non-zero, \ie, we consider anomalous top 
couplings to the photon and $Z$ separately. The results of this analysis 
are shown in Figs.3a-b and Figs. 4a-b which compare a 500 
and 1000 GeV NLC at two different 
values of the integrated luminosity. As above, for the 500(1000) GeV case, 
the minimum 
gluon jet energy, $E_g^{min}$, was set to 25(50) GeV corresponding to 
$z\geq 0.1$. A fixed energy bin width of $\Delta z$=0.05 was chosen in both 
cases so that at 500(1000) GeV 8(15) bins were used to cover the entire 
spectrum. In either case the constraints on the anomalous $t\bar t\gamma$ 
couplings are seen to be stronger qualitatively than the corresponding 
$t\bar tZ$ ones. In 
the former case the allowed region is essentially a long narrow circular band, 
which has been cut off at the top for the $\sqrt s=$500 GeV NLC. In the 
latter case, the allowed 
region lies inside a rather large ellipse. We see from these figures that to 
increase the sensitivity to anomalous couplings it is far better to go to 
higher center of mass 
energies than to simply double the statistics of the sample. The 1 TeV results 
are seen to be significantly better than those quoted in Ref.{\cite {old}} due 
to lower value of $E_g^{min}$. Note that with data from only a single center 
of mass energy available there can be some ambiguity in the values of the 
anomalous couplings. By combining data from two different energies such 
ambiguities can be completely removed and the size of the allowed region 
shrinks drastically, as can be seen in Fig.5.

\vspace*{-0.5cm}
\nn
\begin{figure}[htbp]
\centerline{
\psfig{figure=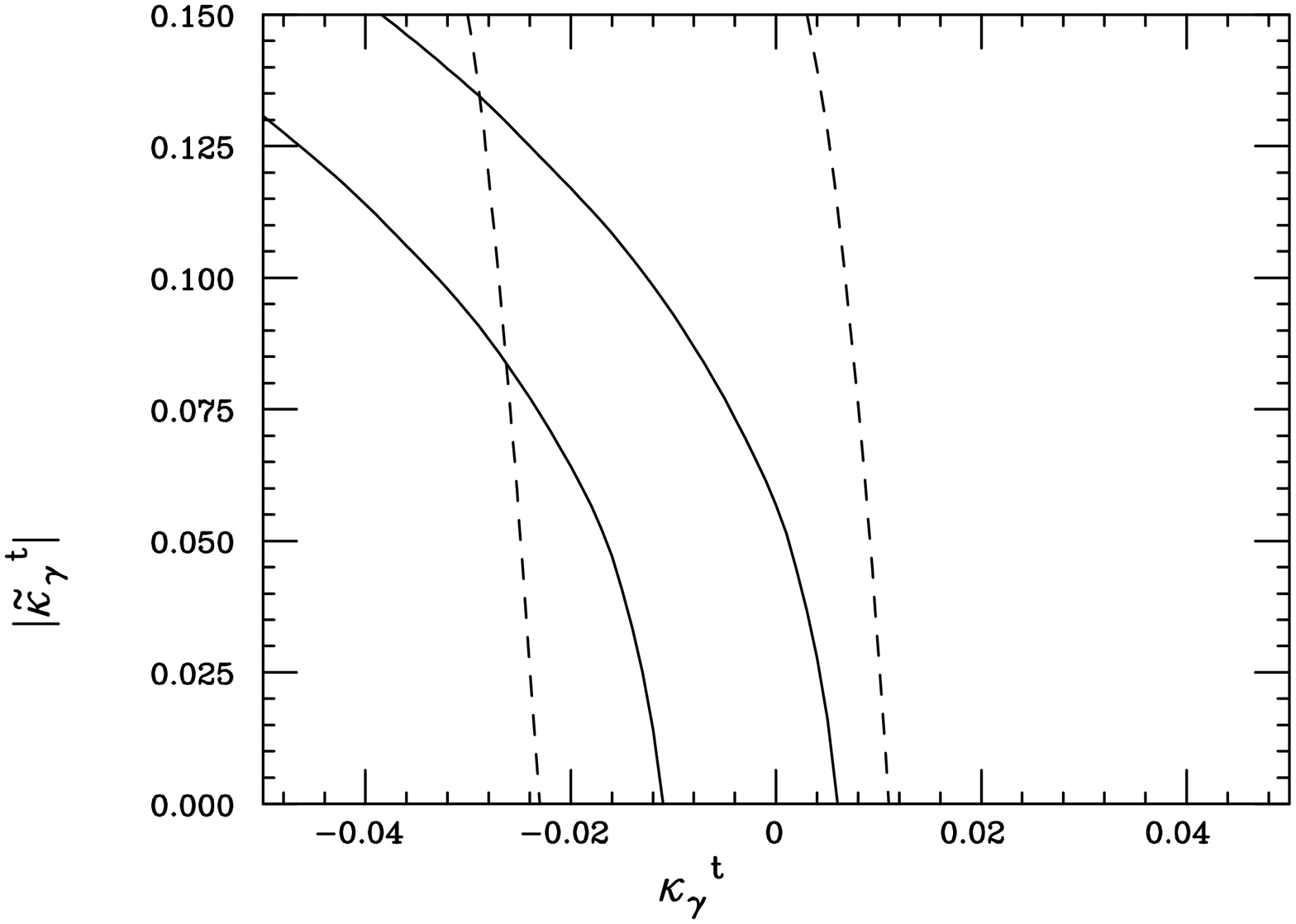,height=9.1cm,width=9.1cm,angle=-90}
\hspace*{-5mm}
\psfig{figure=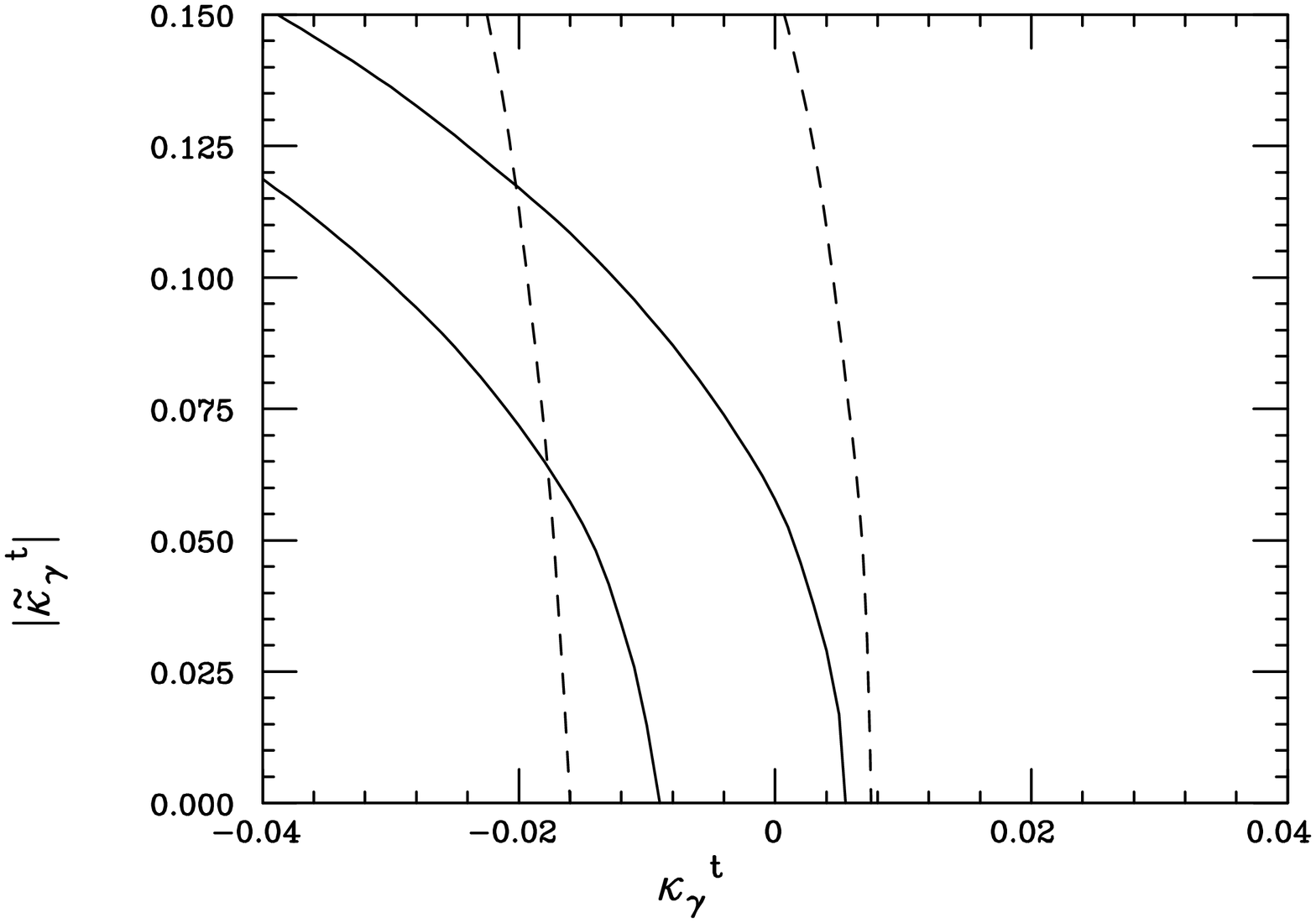,height=9.1cm,width=9.1cm,angle=-90}}
\vspace*{-1cm}
\caption{\small Expanded view of the overlapping regions from Fig.3a(4a) shown 
on the left(right).}
\end{figure}
\vspace*{0.4mm}

\section{Discussion and Conclusions}

In this paper we have shown that the 
process $e^+e^- \to t\bar tg$ can be used to obtain stringent limits on the 
anomalous dipole-like couplings of the top to $\gamma,g$ and $Z$ through 
an examination of the associated gluon energy spectrum. Such 
measurements are seen to be complementary to those which directly probe the 
$t\bar t$ production vertex at hadron and $e^+e^-$ colliders. By combining 
both sets of data a very high sensitivity to the anomalous couplings can 
be achieved. The 
value of the cut on the gluon energy was shown to play a key role in obtaining 
strong bounds on these anomalous couplings. These results may be strengthened 
in the future if we find that the $E_g^{min}$ cut can be further softened.

\Acknowledgements

The author would like to thank P. Burrows, D. Atwood, A. Kagan and 
J.L. Hewett for discussions related to this work.

%
\def\MPL #1 #2 #3 {Mod.~Phys.~Lett.~{\bf#1},\ #2 (#3)}
\def\NPB #1 #2 #3 {Nucl.~Phys.~{\bf#1},\ #2 (#3)}
\def\PLB #1 #2 #3 {Phys.~Lett.~{\bf#1},\ #2 (#3)}
\def\PR #1 #2 #3 {Phys.~Rep.~{\bf#1},\ #2 (#3)}
\def\PRD #1 #2 #3 {Phys.~Rev.~{\bf#1},\ #2 (#3)}
\def\PRL #1 #2 #3 {Phys.~Rev.~Lett.~{\bf#1},\ #2 (#3)}
\def\RMP #1 #2 #3 {Rev.~Mod.~Phys.~{\bf#1},\ #2 (#3)}
\def\ZP #1 #2 #3 {Z.~Phys.~{\bf#1},\ #2 (#3)}
\def\IJMP #1 #2 #3 {Int.~J.~Mod.~Phys.~{\bf#1},\ #2 (#3)}

\end{document}